\documentclass[sigconf]{acmart}
\usepackage{xspace}
\usepackage{longtable}
\usepackage{array}
\usepackage[table]{xcolor}

\usepackage{color}
\usepackage{nameref}
\usepackage{multirow}
\usepackage{tabularx} % for automatic sizing of tables
\usepackage{censor}
\usepackage{nameref}
\usepackage{enumitem}
\usepackage{ragged2e} % for \RaggedRight in tabularx

\usepackage[most]{tcolorbox}

\usepackage{xcolor}          % For custom colors

\definecolor{lightblue}{RGB}{0, 0, 100}

\newtcolorbox{MyBox}{
  colback=white,
  colframe=lightblue,
  fonttitle=\bfseries,
  coltitle=black,
  sharp corners,
  boxrule=1pt,
  left=5pt,
  right=5pt,
  top=5pt,
  bottom=5pt,
  breakable
}

\usepackage{xcolor}
\usepackage{mdframed}

\newmdenv[
  backgroundcolor=black!6,
  leftline=true, rightline=false, topline=false, bottomline=false,
  linecolor=black, linewidth=3pt,
  innerleftmargin=10pt, innerrightmargin=10pt,
  innertopmargin=8pt, innerbottommargin=8pt,
  skipabove=8pt, skipbelow=8pt
]{rqbox}

\newcommand{\RQbox}[2]{%
\begin{rqbox}
\textbf{#1}\par\medskip
\textit{#2}
\end{rqbox}
}

\usepackage[most]{tcolorbox}
\usepackage{xcolor}

\newtcolorbox{databox}{
  colback=teal!20!white,   % background color
  boxrule=0pt,             % no outer border
  enhanced,
  borderline west={4pt}{0pt}{black}, % thick left bar
  left=10pt,
  right=10pt,
  top=8pt,
  bottom=8pt,
  sharp corners
}

\usepackage{fontawesome5}

\newtcolorbox{resultbox}{
  enhanced,
  colback=blue!5,
  boxrule=0pt,
  sharp corners,
  borderline west={4pt}{0pt}{black},
  left=10pt,
  right=10pt,
  top=8pt,
  bottom=8pt
}

\AtBeginDocument{%
  }

\setcopyright{acmlicensed}
\copyrightyear{2016}
\acmYear{2016}
\acmDOI{TBDXXXX.XXXXXXX}
\acmBooktitle{Journal Ahead Workshop (JAWs) - 41st IEEE/ACM International Conference on Automated Software Engineering}
\acmISBN{978-1-4503-XXXX-X/2018/06}

\begin{document}

\title{Technostress in the Age of AI: A Preliminary Study with Software Professionals}


\author{Ronnie de Souza Santos}
\email{ronnie.desouzasantos@ucalgary.ca}
\affiliation{%
  \institution{University of Calgary}
  \city{Calgary}
  \state{Alberta}
  \country{Canada}
}

\author{Italo Santos}
\email{isantos3@hawaii.edu}
\affiliation{%
  \institution{University of Hawai‘i at Mānoa}
  \city{Honolulu}
  \state{Hawaii}
  \country{USA}
  }

\author{Cleyton Magalhaes}
\email{cleyton.vanut@ufrpe.br}
\affiliation{%
  \institution{UFRPE}
  \city{Recife}
  \state{Pernambuco}
  \country{Brazil}}

\author{Zixuan Feng}
\affiliation{%
  \institution{Virginia Commonwealth University}
  \city{Richmond}
  \state{Virginia}
  \country{USA}
}
\email{fengz3@vcu.edu}

\begin{abstract}
The rapid adoption and evolution of AI are changing software engineering work and requiring professionals to repeatedly adapt their knowledge, practices, and skills. Although technological adaptation has long characterized software development, less is known about how these new and recurring pressures manifest as technostress. This preliminary exploratory study investigates AI related technostress among software professionals. We conducted a survey and performed a thematic analysis of responses from 121 software professionals across 26 countries reporting their recent experiences with AI at work. Our findings suggest that AI related technostress emerges not only from adapting to rapidly changing technologies, but also from having to manage the work, technical responsibilities, and professional changes that accompany their adoption. This characterization shows that AI introduces pressures beyond learning and using new tools, affecting how software professionals perform and remain accountable for technical work and how they prepare for the future of their careers.
\end{abstract}


%\received{20 October 2025}


\maketitle

\section{Introduction}
\label{sec:introduction}

AI is becoming part of software development practice, supporting activities across the software lifecycle and changing how software professionals perform technical work~\cite{jeyam2026still, anwar2025software, farrag2026productivity}. With the adoption of Generative AI (GenAI) and Large Language Model (LLM) based tools, some development activities are shifting from producing software artifacts toward evaluating, reviewing, and supervising AI generated work~\cite{farrag2026productivity}. These changes introduce new demands related to verification, technical judgment, productivity, and the competencies required to work with AI~\cite{anwar2025software, farrag2026productivity}.

Although adaptation to technological change has long been part of software engineering, GenAI introduces additional and recurring pressures. Software professionals must keep pace with changing AI capabilities, determine when generated outputs can be trusted, develop new competencies, and respond to changing expectations regarding productivity and professional roles~\cite{kwon2026investigating, kumar2024machine, sari2026ai, wong2025s}. These demands can be understood through \textit{technostress}, which describes strain associated with the use of and adaptation to technologies~\cite{kwon2026investigating, kumar2024machine, jeyam2026still}. In the context of AI, technostress has been associated with learning demands, cognitive load, technological uncertainty, changing professional roles, job insecurity, and the need to verify AI generated outputs~\cite{kumar2024machine, sari2026ai, kwon2026investigating}.

Previous software engineering research has investigated AI adoption, use, productivity, trust, skills, and changes to development practices~\cite{anwar2025software, farrag2026productivity, nguyen2025generative, durrani2024decade, fan2023large}. However, less is known about how the pressures created by AI use and repeated adaptation manifest as technostress among software professionals. To address this gap, we investigate how software professionals experience these pressures across their use of AI, their technical work and decision making, and their perceptions of skills, job security, and professional careers. Accordingly, our study is guided by the following research question (RQ):

\RQbox{RQ. Technostress AI and Software Engineering}{How does the growing use of AI and the resulting need for repeated adaptation manifest as technostress among software professionals?}

To answer our RQ, we conducted a survey with software professionals combining open ended questions with quantitative measures of technostress and related professional outcomes. In this paper, we report a preliminary qualitative analysis of the open-ended questions that provides: 1) empirical evidence of how AI use and repeated adaptation create difficulties and pressures in software engineering work; 2) a characterization of AI related technostress across AI use and adaptation, technical work and professional careers; and 3) a qualitative foundation for an extended study that will use the quantitative measures to develop a model of AI related technostress in software engineering.

The remainder of this paper is organized as follows. Section~\ref{sec:background} presents the background of this study. Section~\ref{sec:method} describes the methodology. Section~\ref{sec:findings} reports the findings, which are discussed in Section~\ref{sec:discussion}. Finally, Section~\ref{sec:conclusions} presents the concluding remarks.

\section{Background}
\label{sec:background}

This section discusses how AI, particularly GenAI and LLM based tools, is changing software engineering work and introduces technostress as a lens for understanding the adaptation demands associated with these changes. \\

\noindent\textbf{AI and Changes in Software Engineering Work.}
Currently, software professionals are facing a double edged effect from having AI tools, including GenAI and LLMs, integrated into their workflows. On one hand, these tools can support and automate activities across requirements, design, implementation, testing, maintenance, documentation, and code review, reducing manual effort and potentially increasing productivity~\cite{hou2024large, terragni2025future, nguyen2025generative}. On the other hand, these benefits introduce additional demands and expectations. Emerging discussions suggest that organizational expectations surrounding AI adoption and productivity may encourage professionals to increase visible AI interaction to signal productivity or organizational alignment, a practice described as ``Tokenmaxxing''~\cite{smite2026beyond}. GenAI outputs can also be nondeterministic, sensitive to input parameters, and affected by hallucinations, requiring attention to the dependability and accuracy of generated artifacts~\cite{nguyen2025generative}. Productivity gains can also be accompanied by additional verification and review effort~\cite{farrag2026productivity}. Beyond artifact quality, AI use introduces concerns related to privacy, security, intellectual property, licensing, transparency, and accountability~\cite{sallou2024breaking, ahmed2025artificial}.

Moreover, these demands are not static. AI models and development tools are continually updated, with new models, features, and capabilities becoming available. Changes to models can alter their outputs over time, creating additional uncertainty regarding their use and reproducibility~\cite{sallou2024breaking}. At the same time, the integration of GenAI continues to raise open questions concerning professional competencies, software processes and tools, engineering management, transparency, data accessibility, dependability, and other aspects of its practical adoption~\cite{nguyen2025generative}. This creates an environment in which software professionals may need to repeatedly learn how to use these technologies, reconsider how they fit within existing workflows, and adapt their practices as their capabilities and conditions of use change. Thus, the challenge for software professionals is not only adapting to AI, but continuously adapting as the technology and the practices surrounding its use continue to change. \\

\noindent\textbf{Technostress and Continuous Technological Adaptation.}
The continuous adaptation required by technological change can itself become a source of stress. This phenomenon is commonly conceptualized as technostress, originally described as the difficulty in adapting to new computer technologies in a healthy manner~\cite{xia2023coworking}. Technostress is commonly characterized through five stressors: \textit{techno-overload}, when technology requires individuals to work more or faster; \textit{techno-invasion}, when technology blurs the boundaries between work and personal life; \textit{techno-complexity}, when individuals perceive their technological skills as inadequate and require additional time and effort to understand the technology; \textit{techno-insecurity}, involving fear of job loss due to technological advancements; and \textit{techno-uncertainty}, arising from continuous technological updates and the resulting need for adaptation~\cite{kwon2026investigating, jeyam2026still}.

The growing integration of AI into professional work illustrates these concepts. Research on AI induced technostress identifies technological complexity and uncertainty, reliability, role ambiguity, job insecurity, and the continued need for skill adaptation among the pressures associated with AI adoption~\cite{kumar2024machine, kwon2026investigating}. These pressures can be compounded by characteristics of AI systems themselves. For example, limited transparency and hallucinations have been associated with control anxiety, which can subsequently reduce creative performance in human AI collaboration~\cite{wang2026unravelling}. As these pressures accumulate, they can have consequences for professionals' wellbeing and work. Studies of AI related technostress have reported associations with AI anxiety, job anxiety, anxiety, depression, and reduced quality of life~\cite{chang2024does, zhang2025anxiety, litan2025mental, litan2025impact}.

Today, these pressures are particularly relevant to software professionals, as technostress was already present in software development before the widespread adoption of GenAI. In the past, software developers reported technostress associated with technological complexity, uncertainty, workload, and the need to learn and adapt to rapidly changing technologies~\cite{siitonen2022emergence}. More recently, however, GenAI and LLM based tools have added another layer of adaptation, placing professionals under constant pressure to learn new skills and tools, with consequences for their perceptions of competence and relatedness~\cite{wong2025s}. In this scenario, GenAI has a dual role: it can help professionals manage existing demands through learning support and task automation, while simultaneously creating new demands associated with the pace of technological change, skill obsolescence, job insecurity, and the need to assess generated outputs~\cite{jeyam2026still}.
\section{Methodology} 
\label{sec:method}

We conducted a survey to investigate how AI use and repeated adaptation manifest as technostress among software professionals, following established survey research guidelines~\citep{linaker2015guidelines, ralph2020empirical}. The complete study combines quantitative measures of technostress and related constructs with open ended questions. This paper reports the preliminary qualitative analysis, focusing on AI adaptation, changes in technical work, and concerns regarding skills, job security, and professional futures.

\subsection{Questionnaire Design}
\label{sec:questionnaire}

The questionnaire followed established recommendations for survey research in software engineering, including alignment between the research objectives and survey items and iterative refinement by the research team~\citep{linaker2015guidelines,ralph2020empirical}. Its design was grounded in the Technostress Creators framework~\citep{ragu2008consequences}, which defines five dimensions of technostress: \textit{techno-overload}, \textit{techno-invasion}, \textit{techno-complexity}, \textit{techno-insecurity}, and \textit{techno-uncertainty}. We also drew on Lițan~\citep{litan2025mental}, who adapted these dimensions to AI-related technostress. These studies provided the theoretical basis for adapting the instrument to AI use in professional software development.

We adapted the Technostress Creators items~\citep{ragu2008consequences} to the context of AI use in software development, following the AI-focused contextualization by Lițan~\citep{litan2025mental}. The five original dimensions were retained, with items reworded to explicitly refer to AI technologies and software development. The resulting items addressed increased workload and pace of work (\textit{techno-overload}), intrusion of technology into personal time (\textit{techno-invasion}), difficulties associated with learning and using AI technologies (\textit{techno-complexity}), concerns about employment and professional relevance (\textit{techno-insecurity}), and uncertainty associated with continuous technological changes (\textit{techno-uncertainty}). The questionnaire also included constructs related to the professional and emotional consequences of AI use: \textit{professional self-efficacy}, \textit{AI dependence}, \textit{work and skill quality}, \textit{social comparison and competitive pressure}, \textit{career insecurity}, and \textit{AI-related emotional exhaustion and anxiety}. These constructs addressed professionals' confidence in performing their work, reliance on AI, perceptions of their technical work and skills, competitive pressures, career concerns, and emotional strain associated with AI use.

The final questionnaire was organized into eight sections. The first section covered the five technostress creators. Sections~2--7 covered professional self-efficacy, AI dependence, work and skill quality, social comparison and competitive pressure, career insecurity, and AI-related emotional exhaustion and anxiety. The final section collected demographic and professional information, including country, gender identity, professional role and experience, AI experience and usage, employer-provided AI training, clarity of organizational AI policies, work arrangement, organization size, ethnic or racial background, and identification with groups underrepresented in computing. Screening and attention-check questions were also included to support data quality. Most structured items used five-point Likert scales ranging from 1 (\textit{Strongly disagree}) to 5 (\textit{Strongly agree}), while demographic and professional questions used categorical, ordinal, and multiple-selection formats. The questionnaire was implemented using Qualtrics\footnote{\url{https://www.qualtrics.com}}.

Due to the page limit, the complete questionnaire is provided in the supplementary material. This preliminary analysis focuses on three open ended questions: 1) \textit{Can you describe an experience, if any, in which using or adapting to AI in your software development work made your job more difficult, frustrating, overwhelming, or stressful?}, which is related to \textit{techno-overload}, \textit{techno-complexity}, \textit{techno-invasion}, and \textit{techno-uncertainty}; 2) \textit{Can you describe an experience, if any, in which AI changed how you perform software development tasks, solve technical problems, or make technical decisions in a way that created difficulty, pressure, uncertainty, or frustration?}, which is related to \textit{techno-complexity}, \textit{techno-overload}, and \textit{techno-uncertainty}; and 3) \textit{Can you describe how, if at all, the growth of AI has created concerns, uncertainty, or pressure regarding your future career in software development?}, which is related to \textit{techno-insecurity} and \textit{techno-uncertainty}.

\subsection{Pilot and Questionnaire Refinement}

Because the questionnaire was constructed primarily using pre established measures from prior research, the refinement process focused on their adaptation to the context of AI use in software engineering rather than the development of a new instrument. The researchers iteratively reviewed the questionnaire to refine wording and ensure consistency with the study context. Three software professionals subsequently completed a pilot version to assess question clarity, phrasing, survey flow, and completion time. Minor adjustments were made based on their feedback before data collection, and pilot responses were excluded from the final dataset.

\subsection{Sampling and Recruitment}
\label{sec:sampling}

Participants were recruited using purposive, convenience, and referral chain sampling~\citep{baltes2022sampling}. Three complementary recruitment strategies were used to reach software practitioners with professional experience in software engineering and AI use across different organizational and geographic contexts. First, we used Prolific, an online recruitment platform commonly used in empirical software engineering research that supports participant screening~\citep{reid2022software, russo2022recruiting}. We targeted professionals working in software engineering or related computing activities across all continents. This strategy resulted in 102 responses. Second, we recruited professionals from a large South American software company that has operated since 1996 and employs more than 1,200 professionals, with over 70\% working directly in software development across approximately 50 teams. These teams develop software for clients across North America, Latin America, Europe, and Asia, providing access to practitioners working across different technical and project contexts within the same organization. This strategy resulted in 16 responses. Finally, we used convenience and referral chain sampling through our professional networks~\citep{baltes2022sampling}, resulting in 3 responses. In total, we collected 121 responses.

\subsection{Filtering and Data Quality Control}
\label{sec:filtering}

Additional filtering and data quality procedures were applied to responses collected through Prolific. Unlike participants recruited through our industry partner or professional networks, Prolific participants were identified through platform prescreening criteria, which are primarily self reported and may not accurately represent technical skills or professional experience~\citep{reid2022software, russo2022recruiting, alami2024you}. We therefore independently assessed their eligibility and response quality rather than relying solely on the platform criteria. Eligibility was assessed using participants' demographic and professional responses. We considered their professional role, years of experience, AI experience, frequency of AI use, and the software development activities for which they used AI to confirm that they belonged to the target population. Participants whose responses did not indicate the required professional or AI experience were excluded. Response quality was assessed across the complete questionnaire. We considered inconsistencies between related questions, patterns suggesting inattentive completion, completion times substantially shorter than expected based on the pilot, and the relevance and meaningfulness of open ended responses. Incomplete submissions and responses providing insufficient information were also excluded. A response was removed when these checks provided sufficient evidence that the participant did not meet the eligibility criteria or that the submission was not sufficiently reliable for analysis. These procedures are consistent with established data quality practices in survey based software engineering research~\citep{danilova2021you, alami2024you}.

\subsection{Data Analysis}

For the preliminary analysis reported in this paper, we stopped data collection at the beginning of August 2026 and analyzed the 121 responses obtained at that point. We used reflexive thematic analysis~\citep{terry2017thematic} to identify recurring patterns in how software professionals experience AI related technostress. The analysis followed four stages:

\begin{itemize}

    \item \textbf{Familiarization and core extraction}: Two researchers independently read the responses and extracted the core content relevant to each question, preserving participants' wording and original meaning as closely as possible.

    \item \textbf{Open coding}: The researchers independently coded the extracted content inductively, assigning one or more codes when a response described multiple experiences. Codes were derived from participants' accounts rather than predefined concepts, resulting in more than 700 coded instances across the three questions analysed.

    \item \textbf{Code reconciliation and categorization}: The researchers compared their codes in consensus meetings, reconciled similar codes, refined unclear interpretations, and resolved disagreements. The resulting codes were grouped by conceptual similarity into higher level categories, with the original responses revisited throughout the process to verify the interpretations.

    \item \textbf{Theme development}: The categories were synthesized into themes that provided higher level explanations of how AI use and adaptation made software development work more difficult, frustrating, overwhelming, or stressful. The themes were developed inductively rather than from predefined technostress dimensions and iteratively refined against the codes, categories, and original responses until consensus was reached.

\end{itemize}

Shared spreadsheets documented the analysis and coding decisions. Analysis continued until further iterations produced no substantive changes to the coding or thematic structure~\citep{ralph2020empirical}. Data collection continued after the qualitative cutoff to obtain a larger sample for the quantitative component of the extended study.

\subsection{Ethics}

This study was conducted in accordance with the institutional ethics guidelines. Participation was voluntary, and informed consent was obtained from all participants before completing the survey. The questionnaire was anonymous, and no personally identifiable information was collected or stored. Participants were informed that they could withdraw from the study at any time before submitting their responses.

\subsection{Positionality Statement}
The authors have backgrounds in empirical software engineering, human aspects of software development, AI in software engineering, and software development in academic and industry contexts. These experiences informed the study but may also influence the interpretation of participants' accounts of AI related technostress. To reduce the influence of individual interpretations, two researchers independently coded the data, discussed disagreements through consensus meetings, and collaboratively developed the themes with reference to participants' original responses. We therefore acknowledge the researchers' backgrounds as part of the interpretive process~\citep{de2025integrating}.

\subsection{Threats to Validity}
\label{sec:limit}

Consistent with guidelines for survey based research~\cite{ralph2020empirical, braun2021online, melegati2024qualitative}, some limitations inherent to the method should be considered when interpreting our findings. \textbf{1. Construct and internal validity}. This study relies on self reported perceptions, which may be affected by recall, social desirability, and differences in question interpretation. To reduce these threats, our questionnaire primarily used established measures adapted to the study context and refined through researcher discussions and pilot testing. \textbf{2. Qualitative depth}. Surveys provide less opportunity for in depth qualitative investigation than interviews. However, the qualitative analysis reported here is preliminary and intended to characterize AI related technostress and establish foundations for the complete study. The broader study will integrate the quantitative measures through statistical analysis and model development. \textbf{3. Reliability}. Thematic analysis involves researcher interpretation. To reduce biases, two researchers independently coded the responses and conducted consensus meetings to discuss differences, refine codes, and develop themes. \textbf{4. External validity}. The 121 responses analyzed in this paper do not constitute a representative sample of software professionals in global dimensions. Therefore, our findings should be interpreted as preliminary patterns rather than statistically generalizable conclusions. At this moment, we continue our data collection to obtain the sample required for the statistical analysis and model development of the complete study.
\section{Results}
\label{sec:findings}

This section presents the study results, beginning with participants' demographic and professional characteristics, followed by qualitative findings on difficulties associated with AI use and adaptation, changes in technical work and decision making, and implications for future software development careers.

\subsection{Participant Characteristics}
\label{sec:participant_characteristics}

Our sample consisted of 121 software professionals from 26 countries across Africa, Asia, Europe, North America, South America, and Oceania: Argentina (1/121), Australia (3/121), Brazil (27/121), Canada (10/121), Chile (2/121), Colombia (1/121), Egypt (6/121), France (2/121), Germany (7/121), India (3/121), Ireland (1/121), Italy (2/121), Japan (4/121), Kenya (3/121), Mexico (3/121), the Netherlands (2/121), New Zealand (2/121), Poland (4/121), Portugal (1/121), Slovakia (1/121), South Africa (6/121), Spain (3/121), Sweden (2/121), the United Kingdom (10/121), the United States (14/121), and Viet Nam (1/121).

Most participants were software developers or engineers (102/121), followed by data scientists or machine learning engineers (9/121), engineering managers or technical leads (5/121), QA professionals (2/121), and other roles (3/121). Professional experience ranged from less than one year (8/121), one to two years (24/121), three to five years (43/121), six to ten years (18/121), 11 to 15 years (13/121), to more than 15 years (15/121). AI experience was considerable (67/121), extensive (29/121), some (23/121), or very little (2/121). Most participants used AI professionally several times a day (49/121) or daily (44/121), while others used it a few times a week (22/121), a few times a month (5/121), or less than once a month (1/121).

Formal employer training on responsible AI use was reported by 62/121 participants, while 55/121 had received none and 4/121 were unsure. AI policies were extremely clear (24/121), very clear (38/121), moderately clear (30/121), slightly clear (15/121), or not at all clear (6/121), while 8/121 reported having no organizational AI policies. Organization sizes ranged from 1 to 9 (10/121), 10 to 49 (20/121), 50 to 249 (21/121), 250 to 999 (17/121), 1,000 to 4,999 (25/121), to 5,000 or more employees (27/121), with 1/121 unsure. Participants worked in hybrid (57/121), primarily remote (36/121), or fully remote arrangements (28/121).

Regarding gender, 80/121 participants identified as men, 40/121 as women, and 1/121 preferred not to disclose. In addition, 59/121 identified with at least one group underrepresented in software engineering or computing, including women, LGBTQIA+ people, neurodivergent people, people of colour, people with disabilities, and people from low income backgrounds; 62/121 indicated that this question was not applicable. These categories were not mutually exclusive.

\subsection{AI as a Source of Difficulty and Stress in Software Development}
\label{sec:ai_use_adaptation}

Four themes characterize the difficulties reported by participants: \textit{AI creates additional work and intensifies work demands}; \textit{AI makes software work harder to understand, control, and maintain}; \textit{AI use and adaptation introduce new sources of strain and constraint}; and \textit{AI changes workplace expectations, responsibilities, and interactions}. A fifth theme, \textit{No difficult AI experience reported}, represents participants who did not report such difficulties or described AI as making their work easier. The distribution of participants across these themes is presented below.

\begin{resultbox}
\label{res:ai_use_adaptation}
\footnotesize
\faChartBar\ \textbf{Overview: Difficulties and stress associated with AI use and adaptation}

\begin{tabularx}{\columnwidth}{Xc}
\hline
\textbf{Theme} & \textbf{Participants} \\
\hline
AI creates additional work and intensifies work demands & 59 \\
AI makes software work harder to understand, control, and maintain & 63 \\
AI use and adaptation introduce new sources of strain and constraint & 29 \\
AI changes workplace expectations, responsibilities, and interactions & 10 \\
No difficult AI experience reported & 13 \\
\hline
\end{tabularx}
\end{resultbox}

\noindent\textbf{AI creates additional work and intensifies work demands.}
For many participants, the difficulty associated with AI came from the additional work required to assess and complete AI assisted tasks. Faster generation was frequently followed by verification, correction, debugging, refinement, and review, sometimes requiring more effort than completing the task without AI. P061 explained that \textit{``Reviewing AI generated code that looked correct but contained subtle hallucinations and inefficient queries actually took longer than writing the feature from scratch.''} At the same time, perceived productivity gains could translate into expectations for greater output and shorter timelines. P032 reported that \textit{``now my bosses think I can finish everything much faster, so they're giving me more work to do in less time.''} AI therefore did not necessarily reduce work demands, but could redistribute effort toward verification and rework while increasing expectations regarding the amount and pace of work.

\noindent\textbf{AI makes software work harder to understand, control, and maintain.}
Difficulties also emerged from the characteristics of the software produced or modified with AI. Participants encountered outputs that appeared plausible but contained errors, did not fit existing architectures or requirements, introduced unnecessary complexity, or were difficult to understand and maintain. P013 described how an \textit{``AI tool generated a code which looked clear and with no errors but later found out it was full of logical errors.''} Beyond correctness, participants described difficulty retaining sufficient understanding of artifacts they remained responsible for. P088 explained that \textit{``you don't have a real deep understanding of what the code does as you didn't write it itself.''} These accounts show how AI could make software work harder by introducing uncertainty about what generated code does, whether it fits the surrounding system, and whether developers can confidently control and maintain it.

\noindent\textbf{AI use and adaptation introduce new sources of strain and constraint.}
Another pattern involved the continuing effort required to adapt to AI while working within the limitations of the available tools. Participants described having to learn changing tools and workflows, keep pace with new practices, and manage limitations in model capabilities and access. These demands were frequently accompanied by frustration, fatigue, anxiety, stress, or cognitive overload. P017 explained that \textit{``there are so many new things that we need to learn each and every week. Which becomes frustrating and stressful overall.''} For P016, the strain was associated directly with handling generated output: \textit{``Reviewing and debugging AI-generated code often proved more mentally exhausting than writing it from scratch.''} The reported strain consequently involved both continuous adaptation to AI and the cognitive and practical demands of working with it.

\noindent\textbf{AI changes workplace expectations, responsibilities, and interactions.}
Some difficulties originated less from the AI tools themselves and more from changes in the workplace surrounding their adoption. Participants described altered expectations from organizations and clients, redistribution of responsibilities, changes in opportunities to practice technical skills, and disruptions to collaboration. P025, for example, reported \textit{``having to take over data science work bc of layoffs due to AI.''} The effects could also extend to interactions among professionals. P119 described AI supported brainstorming as sometimes providing \textit{``only trivial replies or word salad, and often killing the discussion and interaction between people, which is then hard to restore.''} In these accounts, AI adoption changed the conditions under which software work was organized, including who performed particular activities, what was expected from developers, and how people worked with one another.

\noindent\textbf{No difficult AI experience reported.}
Not all participants associated AI use or adaptation with greater difficulty or stress. Some explicitly stated that they had not experienced the situation described in the question, while others reported that AI made their work easier. P030 stated, \textit{``I never had that experience that adapting AI would make my job more difficult.''} P108 described a more explicitly positive experience: \textit{``I haven't had any such experience. On the contrary, AI has been incredibly helpful and has actually made my work much easier.''} This contrasting pattern indicates that experiences were not uniform across the sample, as some participants perceived AI as neutral or beneficial rather than as an additional source of difficulty.

\subsection{How AI Changes Technical Work in Software Development}
\label{sec:technical_work_decisions}

Four themes characterize how AI changes software development tasks, technical problem solving, and technical decision making: \textit{AI shifts technical work from implementation toward verification and correction}; \textit{AI makes technical decision making more uncertain and cognitively demanding}; \textit{AI reduces ownership and control over technical work while responsibility remains with developers}; and \textit{AI accelerates technical work while increasing expectations and weakening opportunities for independent technical development}. A contrasting pattern included participants who reported no relevant difficulty or described AI as facilitating their technical work. The distribution of participants across the themes is presented below.

\begin{resultbox}
\label{res:technical_work_decisions}
\footnotesize
\faChartBar\ \textbf{Overview: How AI changes technical work in software development}

\begin{tabularx}{\columnwidth}{Xc}
\hline
\textbf{Theme} & \textbf{Participants} \\
\hline
AI shifts technical work from implementation toward verification and correction & 58 \\
AI makes technical decision making more uncertain and cognitively demanding & 36 \\
AI reduces ownership and control over technical work while responsibility remains with developers & 22 \\
AI accelerates technical work while increasing expectations and weakening opportunities for independent technical development & 28 \\
\hline
\end{tabularx}
\end{resultbox}

\noindent\textbf{AI shifts technical work from implementation toward verification and correction.}
A common change involved the redistribution of technical effort from producing solutions to assessing and correcting what AI produced. Participants described reviewing, testing, validating, debugging, and adapting generated code when outputs contained errors, failed to meet requirements, or did not fit the surrounding system. P003 reported that although AI generated work quickly, \textit{``it took a lot of work to go back through and ensure it met the team's standards and used best practices.''} For some, this amounted to a change in the nature of development work itself. P016 described \textit{``shifting my primary role from writing original code to critically auditing and filtering AI outputs.''} AI could therefore reduce direct implementation effort while creating a corresponding need for human verification and correction.

\noindent\textbf{AI makes technical decision making more uncertain and cognitively demanding.}
Rather than always simplifying technical decisions, AI sometimes introduced competing alternatives and additional information that developers had to assess. P002 noted that \textit{``different AI tools often provide conflicting suggestions as to the best method,''} requiring the developer to determine which recommendation should be followed. The abundance of alternatives could itself increase the effort required to reach a decision. P115 explained that \textit{``the choices it presents grow exponentially,''} making technical decisions harder and requiring additional analysis, testing, and verification. Across these accounts, AI expanded the set of possible solutions while leaving developers responsible for determining which solution was technically appropriate.

\noindent\textbf{AI reduces ownership and control over technical work while responsibility remains with developers.}
Another change was related to developers' relationship with software generated or modified by AI. Participants described having less knowledge of generated code, reduced ownership of technical artifacts, and less control over AI initiated changes, even though they remained responsible for the resulting software. P022 associated the shift toward reviewing rather than writing code with decreasing code ownership and warned that developers could become \textit{``unable to explain why something works the way it does.''} This tension was stated directly by P024, who explained that developers were \textit{``still responsible for the consequences of that code, but you have way less control over it.''} AI involvement could therefore distance developers from the production and underlying logic of software without transferring responsibility for its quality or consequences.

\noindent\textbf{AI accelerates technical work while increasing expectations and weakening opportunities for independent technical development.}
Faster development was accompanied in some accounts by greater expectations for output and less time for learning and independent problem solving. P066 reported that \textit{``Management now expects us to deliver features way faster because 'we have AI now'.''} Under these conditions, participants described shorter timelines, pressure to produce more, and reduced opportunities to develop technical understanding. P098 stated, \textit{``I feel pressured to output more because of AI and I don't have enough time to learn new concepts in depth.''} Others questioned whether reliance on AI could weaken independent problem solving or basic coding skills. The acceleration associated with AI could therefore affect not only delivery expectations, but also the time and opportunities available for professionals to develop and maintain technical expertise.

\noindent\textbf{Experiences without reported technical difficulty.}
Not all participants associated AI related changes in technical work with difficulty, uncertainty, or frustration. Some reported no relevant experience, while others described reductions in time and effort. P006, for example, reported that using AI for continuous log analysis \textit{``reduced a lot of manual operation,''} while P121 described AI as a source of assistance when difficulty, pressure, uncertainty, or frustration arose. These experiences provide a contrasting pattern in which AI facilitated technical work rather than introducing the difficulties reported in the four themes above.

\subsection{Career Concerns and Uncertainty Associated with AI}
\label{sec:ai_career}

Six themes characterize how the growth of AI shapes concerns, uncertainty, and pressure regarding future careers in software development: \textit{AI creates insecurity about the viability of software development careers}; \textit{AI disrupts entry and progression pathways into software development}; \textit{AI changes the roles and responsibilities of software developers}; \textit{AI creates uncertainty about the value and development of professional expertise}; \textit{AI creates pressure for continuous career adaptation and repositioning}; and \textit{AI creates new organizational pressures and expectations for software developers}. A contrasting pattern involved participants who perceived AI as an opportunity rather than a career threat.

\begin{resultbox}
\label{res:career_uncertainty_pressure}
\footnotesize
\faChartBar\ \textbf{Overview: Career concerns, uncertainty, and pressure}

\begin{tabularx}{\columnwidth}{Xc}
\hline
\textbf{Theme} & \textbf{Participants} \\
\hline
AI creates insecurity about the viability of software development careers & 55 \\
AI changes the roles and responsibilities of software developers & 36 \\
AI creates pressure for continuous career adaptation and repositioning & 31 \\
AI is perceived by some developers as an opportunity rather than a career threat & 29 \\
AI creates new organizational pressures and expectations for software developers & 25 \\
AI creates uncertainty about the value and development of professional expertise & 24 \\
AI disrupts entry and progression pathways into software development & 21 \\
\hline
\end{tabularx}

\end{resultbox}

\noindent\textbf{AI creates insecurity about the viability of software development careers.}
Participants associated AI with declining demand for developers, layoffs, and greater difficulty finding and retaining employment. P001 described \textit{``uncertainty and anxiety for landing the next job because of the lack of positions,''} while others feared that fewer developers would be needed as AI capabilities increased. These concerns positioned AI as a source of uncertainty about the continued viability and stability of software development careers.

\noindent\textbf{AI disrupts entry and progression pathways into software development.}
Concerns were particularly pronounced around junior and early career positions, where participants perceived AI as increasingly performing work previously assigned to less experienced developers. P017 reported that companies were \textit{``not hiring as many juniors as they used to,''} while others worried that fewer routine tasks would also reduce opportunities to develop foundational knowledge. AI was therefore perceived as affecting both entry into software development and the opportunities through which new developers gain experience.

\noindent\textbf{AI changes the roles and responsibilities of software developers.}
Participants anticipated a shift from routine implementation toward architecture, technical decision making, and supervision of AI generated work. P066 worried that development could change from \textit{``actual coding and problem solving to just reviewing AI code all day.''} Career uncertainty consequently concerned not only whether developer jobs would remain, but also what those jobs would involve.

\noindent\textbf{AI creates uncertainty about the value and development of professional expertise.}
Another recurring concern involved which technical skills would remain valuable as AI capabilities expanded. P004 described uncertainty about \textit{``the skills I need to develop,''} while others worried about programming skill obsolescence or the erosion of expertise through AI use. Participants therefore questioned both the future value of existing skills and the expertise that future software development would require.

\noindent\textbf{AI creates pressure for continuous career adaptation and repositioning.}
Uncertainty about future skills translated into pressure to continuously learn, adapt, and reconsider professional direction. P003 described an increased need for \textit{``continuous learning and staying up to date with technological advancements,''} while others reported moving toward AI related, architectural, strategic, or verification activities. Remaining employable was consequently associated with continuing adaptation as AI changes software development work.

\noindent\textbf{AI creates new organizational pressures and expectations for software developers.}
Participants also connected AI with expectations for greater productivity, faster delivery, increased AI use, and reduced staffing. P076 described demands for \textit{``quality work in minimal time,''} while still being responsible when faster delivery resulted in problems. Career pressure therefore arose partly from how organizations interpreted and implemented AI capabilities in decisions about performance and work.

\noindent\textbf{AI is perceived by some developers as an opportunity rather than a career threat.}
Not all participants reported career concerns. Some viewed AI as a useful tool that could not replace important aspects of software development, while others associated it with new professional opportunities. P108, for example, reported no concerns and described AI as helping their career, providing a contrasting experience in which AI supported rather than threatened career development.

\RQbox{RQ. Technostress AI and Software Engineering}{The introduction of new AI tools and processes into software development and the repeated adaptation they require manifest as technostress through additional verification and correction work, continuous learning, cognitive overload, frustration, and pressure to work faster. Technical work becomes more focused on evaluating AI generated solutions and making decisions about outputs developers may not fully understand or control, while responsibility for software quality remains with them. These changes also create pressure around keeping skills relevant, adapting to changing roles, finding and maintaining employment, progressing from junior to more advanced roles, and determining how software development careers will evolve.}
\section{Discussion} \label{sec:discussion}

In this section, we discuss our findings in relation to the existing literature on technostress and consider their implications for software development research and practice.

\subsection{Technostress in Software Engineering Following Recent AI Adoption}
\label{sec:discussion_technostress}

Technostress was already present in software development through workload, complexity, and continuous technological change~\cite{siitonen2022emergence}. In other domains, recent literature shows that AI adds pressures related to adaptation, unreliable outputs, reduced control, job insecurity, and changing professional roles~\cite{kumar2024machine,jeyam2026still,kwon2026investigating,sari2026ai}. Our study identifies similar pressures in software engineering across AI use, technical work, and future careers.

\subsubsection{How AI Related Technostress Manifests in Software Development}
 \label{sec:discussion_ai_technostress}
Participants' experiences with AI use and adaptation in software development align with recent findings on AI related technostress, including continuous learning, cognitive demands, unreliable or difficult to interpret outputs, uncertainty, and reduced control~\cite{kumar2024machine,jeyam2026still,kwon2026investigating,sari2026ai,wang2026unravelling}. This is reflected in the themes around \textbf{general AI use and adaptation} in software workflows, which show that AI adds work, intensifies demands, and makes software harder to understand, control, and maintain. The \textbf{technical work} themes further show a shift from direct implementation toward evaluating, verifying, correcting, and understanding AI generated code and solutions. This redistribution creates additional demands, as developers must judge correctness, understand code they did not write, and fix AI introduced problems. Reduced involvement in implementation can also weaken developers' sense of ownership over the software even as responsibility for its quality remains theirs, creating uncertainty, frustration, and cognitive pressure in technical decision making. These findings align with evidence that technostress can arise from unreliable or difficult to interpret outputs, complexity, and reduced control, requiring continued human evaluation of AI generated code~\cite{kumar2024machine,jeyam2026still,kwon2026investigating,sari2026ai,wang2026unravelling}. Finally, career development and work expectations in software engineering also reflect known forms of AI related technostress, including job insecurity from possible technological replacement~\cite{kumar2024machine,xia2023coworking}, pressure to continuously learn and adapt technical skills~\cite{kumar2024machine,sari2026ai}, and challenges to developers' professional expertise and identity as automation expands~\cite{kwon2026investigating}. The \textbf{career related themes} extend these pressures to the broader structure of software development careers, suggesting AI may affect not only job security but also how junior developers enter the profession and build the technical expertise needed to advance.

\subsubsection{AI Related Technostress Dimensions in Software Engineering}
\label{sec:discussion_technostress_dimensions}

The themes that emerged from our analysis reflect four established dimensions of technostress. \textbf{Techno-overload} appears in the additional work required to verify and correct AI generated outputs and in expectations to produce more work faster, while \textbf{techno-complexity} is reflected in difficulties understanding generated solutions, evaluating their correctness, and developing the knowledge required to use AI effectively. \textbf{Techno-uncertainty} is visible in continuous changes to AI tools, practices, required skills, and developer roles. \textbf{Techno-insecurity} is reflected in uncertainty about job availability and the continued viability of software development positions, while the themes around career progression and the future value of professional expertise suggest that insecurity associated with AI may extend beyond the threat of losing an existing job. These dimensions are consistent with recent AI related technostress research, where overload, complexity, uncertainty, and insecurity have been repeatedly identified~\cite{kumar2024machine,sari2025mapping,sari2026ai}, and with recent evidence from software engineering showing complexity, overload, and uncertainty as frequent technostressors and increased insecurity following GenAI adoption~\cite{jeyam2026still}. They also show continuity with technostress documented in software development before widespread GenAI adoption~\cite{siitonen2022emergence}. In contrast, \textbf{techno-invasion} was not identified in the themes emerging from our analysis so far, although it has been reported in studies measuring AI related technostress in other contexts~\cite{litan2025mental,litan2025impact}.

\subsubsection{Technostress and Positive Experiences with AI} \label{sec:discussion_positive_ai}
Although technostress emphasizes the difficulties associated with technology use, the themes also show that AI was not experienced negatively by all software professionals. Some software professionals are experience AI a facilitator of technical work more than a challenge, through faster task completion, support for problem solving and learning, and reduced effort, while others perceived its growth as creating opportunities for career development rather than insecurity. This duality is consistent with research showing that AI related demands can be appraised as challenges that support learning, work engagement, and professional growth~\cite{chang2024does,zhang2025anxiety}, as well as evidence that GenAI can act as both a source and a mitigator of technostress~\cite{jeyam2026still}.

\subsection{Implications for Research and Practice}
\label{sec:implications}

Although preliminary, this study provides empirical evidence of how AI related technostress is experienced by software professionals across AI use and adaptation, technical work, and career uncertainty. For research, it provides a qualitative foundation for a broader model integrating technostress creators with emotional exhaustion, professional self efficacy, AI dependence, work and skill quality, competitive pressure, career insecurity, among other aspects of the software engineering work. For practice, it supports professionals and organizations in understanding and addressing the pressures associated with AI adoption and frequent adaptation due to the rapid evolution of AI technologies, including verification demands, developer ownership and responsibility, and the continuous development of technical expertise. Therefore, the novelty of this study lies in characterizing how AI related technostress manifests specifically in software engineering, connecting pressures associated with AI use and adaptation to changes in technical work, professional expertise, and software development careers.
\section{Conclusions and Future Work} 
\label{sec:conclusions}

This study investigated how AI adoption and the continuous adaptation it demands manifest as technostress among software professionals. A preliminary qualitative analysis of the experience of 121 professionals across 26 countries shows that AI related technostress spans daily work, technical practice, and career expectations, manifesting as heightened verification and learning demands, cognitive and productivity pressure, shifts in technical responsibility and control, and ongoing pressure to adapt skills and careers. Our future work will extend this research by increasing the sample size by approximately 75\% to strengthen and refine the qualitative analysis. We will also analyze the survey's quantitative measures using Partial Least Squares Structural Equation Modeling (PLS-SEM) to develop and evaluate a statistical model of AI-related technostress in software engineering. The model will characterize the relationships among technostress creators, emotional exhaustion and anxiety, professional self-efficacy, AI dependence, work and skill quality, social comparison and competitive pressure, and career insecurity. This analysis will allow us to examine both the measurement properties of the constructs and the strength and direction of the hypothesized relationships among them, providing a more comprehensive understanding of how AI-related technostress emerges and affects software professionals.%Our future work focus on extending this research by expanding our sample by 75\% to refine this qualitative analysis, as well as analyze the survey's quantitative measures to build a model characterizing the relationships between technostress creators, emotional exhaustion and anxiety, professional self efficacy, AI dependence, work and skill quality, social comparison and competitive pressure, and career insecurity in software engineering.

\section{Supplementary Material} 
The supplementary material, including the complete questionnaire and qualitative analysis codebook, is available at: \url{https://figshare.com/s/7c74a60251efceb8f662}.

\nocite{*}

\bibliographystyle{ACM-Reference-Format}
\bibliography{bibliography}

@article{anwar2025software,
  title={Software Engineering a Journey Beyond Code},
  author={Anwar, Ashif},
  journal={Journal of Computer Science and Technology Studies},
  volume={7},
  number={4},
  pages={619--627},
  year={2025}
}

@article{litan2025mental,
  title={Mental health in the “era” of artificial intelligence: technostress and the perceived impact on anxiety and depressive disorders—an SEM analysis},
  author={Lițan, Daniela-Elena},
  journal={Frontiers in Psychology},
  volume={16},
  pages={1600013},
  year={2025},
  publisher={Frontiers Media SA}
}

@article{ragu2008consequences,
  title={The consequences of technostress for end users in organizations: Conceptual development and empirical validation},
  author={Ragu-Nathan, TS and Tarafdar, Monideepa and Ragu-Nathan, Bhanu S and Tu, Qiang},
  journal={Information systems research},
  volume={19},
  number={4},
  pages={417--433},
  year={2008},
  publisher={Informs}
}

@inproceedings{coutinho2024role,
  title={The role of generative ai in software development productivity: A pilot case study},
  author={Coutinho, Mariana and Marques, Lorena and Santos, Anderson and Dahia, Marcio and Fran{\c{c}}a, Cesar and de Souza Santos, Ronnie},
  booktitle={Proceedings of the 1st ACM International Conference on AI-Powered Software},
  pages={131--138},
  year={2024}
}

@inproceedings{santos2025model,
  title={Model-assisted and human-guided: Perceptions and practices of software professionals using llms for coding},
  author={Santos, Italo and Magalhaes, Cleyton and Santos, Ronnie De Souza},
  booktitle={2025 2nd IEEE/ACM International Conference on AI-powered Software (AIware)},
  pages={105--112},
  year={2025},
  organization={IEEE}
}

@inproceedings{ullrich2025requirements,
  title={From requirements to code: understanding developer practices in LLM-assisted software engineering},
  author={Ullrich, Jonathan and Koch, Matthias and Vogelsang, Andreas},
  booktitle={2025 IEEE 33rd International Requirements Engineering Conference (RE)},
  pages={257--266},
  year={2025},
  organization={IEEE}
}

@article{khati2025mapping,
  title={Mapping the trust terrain: LLMs in Software Engineering-insights and perspectives},
  author={Khati, Dipin and Liu, Yijin and Palacio, David N and Zhang, Yixuan and Poshyvanyk, Denys},
  journal={ACM Transactions on Software Engineering and Methodology},
  year={2025},
  publisher={ACM New York, NY}
}

@article{zheng2025towards,
  title={Towards an understanding of large language models in software engineering tasks},
  author={Zheng, Zibin and Ning, Kaiwen and Zhong, Qingyuan and Chen, Jiachi and Chen, Wenqing and Guo, Lianghong and Wang, Weicheng and Wang, Yanlin},
  journal={Empirical Software Engineering},
  volume={30},
  number={2},
  pages={50},
  year={2025},
  publisher={Springer}
}

@inproceedings{sallou2024breaking,
  title={Breaking the silence: the threats of using llms in software engineering},
  author={Sallou, June and Durieux, Thomas and Panichella, Annibale},
  booktitle={Proceedings of the 2024 ACM/IEEE 44th International conference on software engineering: new ideas and emerging results},
  pages={102--106},
  year={2024}
}

@article{hou2024large,
  title={Large language models for software engineering: A systematic literature review},
  author={Hou, Xinyi and Zhao, Yanjie and Liu, Yue and Yang, Zhou and Wang, Kailong and Li, Li and Luo, Xiapu and Lo, David and Grundy, John and Wang, Haoyu},
  journal={ACM Transactions on Software Engineering and Methodology},
  volume={33},
  number={8},
  pages={1--79},
  year={2024},
  publisher={ACM New York, NY}
}

@inproceedings{fan2023large,
  title={Large language models for software engineering: Survey and open problems},
  author={Fan, Angela and Gokkaya, Beliz and Harman, Mark and Lyubarskiy, Mitya and Sengupta, Shubho and Yoo, Shin and Zhang, Jie M},
  booktitle={2023 IEEE/ACM International Conference on Software Engineering: Future of Software Engineering (ICSE-FoSE)},
  pages={31--53},
  year={2023},
  organization={IEEE}
}

@article{durrani2024decade,
  title={A decade of progress: A systematic literature review on the integration of AI in software engineering phases and activities (2013-2023)},
  author={Durrani, Usman Khan and Akpinar, Mustafa and Adak, Muhammed Fatih and Kabakus, Abdullah Talha and {\"O}zt{\"u}rk, Muhammed Maruf and Saleh, Mohammed},
  journal={IEEE Access},
  volume={12},
  pages={171185--171204},
  year={2024},
  publisher={IEEE}
}

@article{terragni2025future,
  title={The future of ai-driven software engineering},
  author={Terragni, Valerio and Vella, Annie and Roop, Partha and Blincoe, Kelly},
  journal={ACM Transactions on Software Engineering and Methodology},
  volume={34},
  number={5},
  pages={1--20},
  year={2025},
  publisher={ACM New York, NY}
}

@article{nguyen2025generative,
  title={Generative artificial intelligence for software engineering—A research agenda},
  author={Nguyen-Duc, Anh and Cabrero-Daniel, Beatriz and Przybylek, Adam and Arora, Chetan and Khanna, Dron and Herda, Tomas and Rafiq, Usman and Melegati, Jorge and Guerra, Eduardo and Kemell, Kai-Kristian and others},
  journal={Software: Practice and Experience},
  volume={55},
  number={11},
  pages={1806--1843},
  year={2025},
  publisher={Wiley Online Library}
}

@article{ahmed2025artificial,
  title={Artificial intelligence for software engineering: The journey so far and the road ahead},
  author={Ahmed, Iftekhar and Aleti, Aldeida and Cai, Haipeng and Chatzigeorgiou, Alexander and He, Pinjia and Hu, Xing and Pezz{\`e}, Mauro and Poshyvanyk, Denys and Xia, Xin},
  journal={ACM Transactions on Software Engineering and Methodology},
  volume={34},
  number={5},
  pages={1--27},
  year={2025},
  publisher={ACM New York, NY}
}

@article{farrag2026productivity,
  title={The Productivity-Reliability Paradox: Specification-Driven Governance for AI-Augmented Software Development},
  author={Farrag, Sabry E},
  journal={arXiv preprint arXiv:2605.01160},
  year={2026}
}

@inproceedings{maatta2026generative,
  title={Generative AI and the Future of Professional Software Development: Survey Findings from Finland},
  author={M{\"a}{\"a}tt{\"a}, Samuli and Kelanti, Markus and Turhan, Burak},
  booktitle={Proceedings of the 34th ACM International Conference on the Foundations of Software Engineering},
  pages={1623--1631},
  year={2026}
}

@article{chang2024does,
  title={Does AI-driven technostress promote or hinder employees’ artificial intelligence adoption intention? A moderated mediation model of affective reactions and technical self-efficacy},
  author={Chang, Po-Chien and Zhang, Wenhui and Cai, Qihai and Guo, Hongchi},
  journal={Psychology Research and Behavior Management},
  pages={413--427},
  year={2024},
  publisher={Taylor \& Francis}
}

@article{duong2025unraveling,
  title={Unraveling the dark side of ChatGPT: a moderated mediation model of technology anxiety and technostress},
  author={Duong, Cong Doanh and Ngo, Thi Viet Nga and Khuc, The Anh and Tran, Nhat Minh and Nguyen, Thi Phuong Thu},
  journal={Information Technology \& People},
  volume={38},
  number={4},
  pages={2015--2040},
  year={2025},
  publisher={Emerald Publishing Limited}
}

@article{hogemann2025technostress,
  title={Technostress and generative AI in the workplace: a qualitative analysis of young professionals},
  author={H{\"o}gemann, Malte and Hein, Laura and Britsche, Jan-Oliver and Thomas, Oliver},
  journal={Frontiers in Artificial Intelligence},
  volume={8},
  pages={1728881},
  year={2025},
  publisher={Frontiers Media SA}
}

@article{kumar2024machine,
  title={Machine learning and artificial intelligence-induced technostress in organizations: a study on automation-augmentation paradox with socio-technical systems as coping mechanisms},
  author={Kumar, Amit and Krishnamoorthy, Bala and Bhattacharyya, Som Sekhar},
  journal={International Journal of Organizational Analysis},
  volume={32},
  number={4},
  pages={681--701},
  year={2024},
  publisher={Emerald Publishing Limited}
}

@inproceedings{kwon2026investigating,
  title={Investigating AI-induced Technostress and Coping Strategies of Professionals},
  author={Kwon, Heesung and Oh, Jeesun and Lee, Suyoun and Lee, Sunok and Lee, Sangsu},
  booktitle={Proceedings of the 2026 CHI Conference on Human Factors in Computing Systems},
  pages={1--16},
  year={2026}
}

@article{litan2025impact,
  title={The impact of technostress generated by artificial intelligence on the quality of life: the mediating role of positive and negative affect},
  author={Lițan, Daniela-Elena},
  journal={Behavioral Sciences},
  volume={15},
  number={4},
  pages={552},
  year={2025},
  publisher={MDPI}
}

@inproceedings{sapkota2025technostressors,
  title={From Technostressors to AI-Stressors: A Systematic Literature Review of Stressors Associated with AI Systems},
  author={Sapkota, Pratik and Makkonen, Markus and Pirkkalainen, Henri and Salo, Markus and Topuzovska Latkovikj, Marija and Bednar, Peter and Rajanen, Mikko and K{\"a}vrestad, Joakim and Vallo Hult, Helena and Elbanna, Amany},
  booktitle={CEUR Workshop Proceedings},
  year={2025},
  organization={RWTH Aachen}
}

@article{sari2025mapping,
  title={Mapping and Synthesizing the Landscape of Artificial Intelligence and Technostress: A Hybrid Bibliometric and Systematic Review Approach},
  author={Sari, Indah Mulia and Malek, Mohd Dahlan Hj A and Lestari, Selfiyani and Khoerunnisa, Siti Vania and Hayuningtyas, Krisbandaru},
  journal={Psychological Research and Intervention},
  volume={8},
  number={2},
  pages={56--75},
  year={2025}
}

@article{sari2026ai,
  title={AI Induced Technostress: A Systematic Review of Risks and Opportunities},
  author={Sari, Indah Mulia and Hj, A Malek Mohd Dahlan and Arsyad, Fachry and Rakhman, Gumgum Gumelar Fajar and Lestari, Selfiyani and Khoerunnisa, Siti Vania},
  journal={International Journal of Psychology and Psychological Therapy},
  volume={26},
  number={1},
  pages={5--16},
  year={2026},
  publisher={La Asociaci{\'o}n de An{\'a}lisis del Comportamiento}
}

@article{shehadeh2026role,
  author  = {Shehadeh, Amer and Ibrahim, Qusai and Mehasan, Azza and Jouda, Enas and Shaheen, Khitam},
  title   = {The Role of Interaction with Artificial Intelligence Technologies in Shaping the Psychological Wellbeing of University Students: A Systematic Review and Meta-Analysis},
  journal = {Journal of Daoist Studies},
  year    = {2026},
  volume  = {19},
  number  = {S8},
  pages   = {1451--1447},
  url     = {https://journalofdaoiststudies.org/index.php/journal/article/view/1564}
}

@article{wang2026unravelling,
  title={Unravelling technostress in human--AI collaborative brainstorming: examining the impact of AI technology characteristics on brainstorming outcomes through AI control anxiety},
  author={Wang, Siran and Yan, Qiang and Leng, Jidong},
  journal={Kybernetes},
  pages={1--19},
  year={2026},
  publisher={Emerald Publishing Limited}
}

@inproceedings{xia2023coworking,
  title={Co-working with AI is a double-sword in technostress? An integrative review of human-AI collaboration from a holistic process of technostress},
  author={Xia, Mengting},
  booktitle={SHS Web of Conferences},
  volume={155},
  pages={03022},
  year={2023},
  organization={EDP Sciences}
}

@article{zhang2025anxiety,
  title={Anxiety or engaged? Research on the impact of technostress on employees' innovative behavior in the era of artificial intelligence},
  author={Zhang, Shengtai and Guo, Pengli and Yuan, Yiwei and Ji, Yajun},
  journal={Acta Psychologica},
  volume={259},
  pages={105442},
  year={2025},
  publisher={Elsevier}
}

@article{zhang2025technostress,
  title={The technostress of ChatGPT usage: How do perceived AI characteristics affect user discontinuous usage through AI anxiety and user negative attitudes?},
  author={Zhang, Tingwan and Tong, Qingyan},
  journal={International Journal of Human--Computer Interaction},
  volume={41},
  number={16},
  pages={9918--9929},
  year={2025},
  publisher={Taylor \& Francis}
}

@inproceedings{jeyam2026still,
  title={“Still in the Loop”: Coping with Technostress in DevOps Teams and the Impact of GenAI},
  author={Jeyam, Dharneeka and Wiedemann, Anna and Schwabe, Gerhard and G{\"u}ney, Kadircan},
  booktitle={Proceedings of the IEEE/ACM 48th International Conference on Software Engineering: Software Engineering in Practice},
  pages={590--600},
  year={2026}
}

@inproceedings{wong2025s,
  title={'It's a spectrum': exploring autonomy, competence, and relatedness in software development processes and tools},
  author={Wong, Novia and Cheng, Nai-Yu and Oewel, Bruna and Genuario, Katherine E and Stoeckl, SarahElizabeth and Schueller, Stephen M and Ahmed, Iftekhar and Van Der Hoek, Andr{\'e} and Reddy, Madhu},
  booktitle={Proceedings of the 2025 CHI Conference on Human Factors in Computing Systems},
  pages={1--19},
  year={2025}
}

@inproceedings{smite2026beyond,
  title={Beyond Individual Prompting Efficiency: A Socio-Technical Perspective on Computational Efficiency in AI-Assisted Software Engineering},
  author={Smite, Darja and Wivestad, Viggo Tellefsen and O'Brien, Gabrielle and Santos, Italo and Destefanis, Giuseppe and Baldassarre, Maria Teresa and Lungu, Mircea and Paja, Elda and de Souza Santos, Ronnie},
  booktitle={20th International Symposium on Empirical Software Engineering and Measurement},
  pages={68:1--68:15},
  year={2026}
}

@inproceedings{siitonen2022emergence,
  title={The emergence of technostress in software development work: Technostressors and underlying factors},
  author={Siitonen, Valtteri and Ritonummi, Saima and Salo, Markus and Pirkkalainen, Henri},
  booktitle={CEUR Workshop Proceedings},
  year={2022},
  organization={RWTH Aachen}
}

@article{ralph2020empirical,
  title={Empirical standards for software engineering research},
  author={Ralph, Paul and Ali, Nauman bin and Baltes, Sebastian and Bianculli, Domenico and Diaz, Jessica and Dittrich, Yvonne and Ernst, Neil and Felderer, Michael and Feldt, Robert and Filieri, Antonio and others},
  journal={arXiv preprint arXiv:2010.03525},
  year={2020}
}

@article{linaker2015guidelines,
  title={Guidelines for conducting surveys in software engineering v. 1.1},
  author={Linaker, Johan and Sulaman, Sardar Muhammad and H{\"o}st, Martin and de Mello, Rafael Maiani},
  journal={Lund University},
  volume={50},
  pages={1--64},
  year={2015}
}

@article{seaman1999qualitative,
  title={Qualitative methods in empirical studies of software engineering},
  author={Seaman, Carolyn B.},
  journal={IEEE Transactions on software engineering},
  volume={25},
  number={4},
  pages={557--572},
  year={1999},
  publisher={IEEE}
}

@article{lee1997content,
  title={Content analysis of archival data.},
  author={Lee, Fiona and Peterson, Christopher},
  journal={Journal of Consulting and clinical Psychology},
  volume={65},
  number={6},
  pages={959},
  year={1997},
  publisher={American Psychological Association}
}

@article{russo2022recruiting,
  title={Recruiting software engineers on prolific},
  author={Russo, Daniel},
  journal={arXiv preprint arXiv:2203.14695},
  year={2022}
}

@article{reid2022software,
  title={Software engineering user study recruitment on prolific: An experience report},
  author={Reid, Brittany and Wagner, Markus and d'Amorim, Marcelo and Treude, Christoph},
  journal={arXiv preprint arXiv:2201.05348},
  year={2022}
}

@article{baltes2022sampling,
  title={Sampling in software engineering research: A critical review and guidelines},
  author={Baltes, Sebastian and Ralph, Paul},
  journal={Empirical Software Engineering},
  volume={27},
  number={4},
  pages={94},
  year={2022},
  publisher={Springer}
}

@inproceedings{danilova2021you,
  title={Do you really code? designing and evaluating screening questions for online surveys with programmers},
  author={Danilova, Anastasia and Naiakshina, Alena and Horstmann, Stefan and Smith, Matthew},
  booktitle={2021 IEEE/ACM 43rd International Conference on Software Engineering (ICSE)},
  pages={537--548},
  year={2021},
  organization={IEEE}
}

@inproceedings{alami2024you,
  title={Are you a real software engineer? best practices in online recruitment for software engineering studies},
  author={Alami, Adam and Zahedi, Mansooreh and Ernst, Neil},
  booktitle={Proceedings of the 1st IEEE/ACM International Workshop on Methodological Issues with Empirical Studies in Software Engineering},
  pages={52--57},
  year={2024}
}

@article{terry2017thematic,
  title={Thematic analysis},
  author={Terry, Gareth and Hayfield, Nikki and Clarke, Victoria and Braun, Virginia and others},
  journal={The SAGE handbook of qualitative research in psychology},
  volume={2},
  number={17-37},
  pages={25},
  year={2017},
  publisher={SAGE Publications Ltd}
}

@inproceedings{de2025integrating,
  title={Integrating Positionality Statements in Empirical Software Engineering Research},
  author={De Sousa, Breno Felix and de Souza Santos, Ronnie and Gama, Kiev},
  booktitle={2025 IEEE/ACM International Workshop on Methodological Issues with Empirical Studies in Software Engineering (WSESE)},
  pages={28--35},
  year={2025},
  organization={IEEE}
}

@article{braun2021online,
  title={The online survey as a qualitative research tool},
  author={Braun, Virginia and Clarke, Victoria and Boulton, Elicia and Davey, Louise and McEvoy, Charlotte},
  journal={International journal of social research methodology},
  volume={24},
  number={6},
  pages={641--654},
  year={2021},
  publisher={Taylor \& Francis}
}

@article{melegati2024qualitative,
  title={Qualitative surveys in software engineering research: definition, critical review, and guidelines},
  author={Melegati, Jorge and Conboy, Kieran and Graziotin, Daniel},
  journal={IEEE Transactions on Software Engineering},
  volume={50},
  number={12},
  pages={3172--3187},
  year={2024},
  publisher={IEEE}
}

\appendix

\end{document}